\documentclass[conference]{IEEEtran}
\IEEEoverridecommandlockouts
\usepackage{cite}
\usepackage{amsmath,amssymb,amsfonts}
\usepackage{algorithmic}
\usepackage{graphicx}
\usepackage{textcomp}
\usepackage{xcolor}
\usepackage{format}
\usepackage{multirow}

\newcommand{\soa}{state-of-the-art }

\def\BibTeX{{\rm B\kern-.05em{\sc i\kern-.025em b}\kern-.08em
    T\kern-.1667em\lower.7ex\hbox{E}\kern-.125emX}}

\begin{document}

\title{CDLNet: Robust and Interpretable Denoising through Deep Convolutional Dictionary Learning}

\author{\IEEEauthorblockN{Nikola Janju\v{s}evi\'{c}, Amirhossein Khalilian-Gourtani, Yao Wang}
\IEEEauthorblockA{\textit{Electrical and Computer Engineering Department, New York University,
Brooklyn, USA.} \\
email:\{npj226, akg404, yaowang\}@nyu.edu}
}

\maketitle
\begin{abstract}
Deep learning based methods hold \soa results in image denoising, but remain difficult to interpret due to their construction from poorly understood building blocks such as batch-normalization, residual learning, and feature domain processing. Unrolled optimization networks propose an interpretable alternative to constructing deep neural networks by deriving their architecture from classical iterative optimization methods, without use of tricks from the standard deep learning tool-box. So far, such methods have demonstrated performance close to that of \soa models while using their interpretable construction to achieve a comparably low learned parameter count. In this work, we propose an unrolled convolutional dictionary learning network (CDLNet) and demonstrate its competitive denoising performance in both low and high parameter count regimes. Specifically, we show that the proposed model outperforms the \soa denoising models when scaled to similar parameter count. In addition, we leverage the model's interpretable construction to propose an augmentation of the network's thresholds that enables \soa blind denoising performance and near-perfect generalization on noise-levels unseen during training.
\end{abstract}


\section{Introduction and Background}
In recent years, deep-learning methods have become \soa in imaging inverse
problems such as denoising \cite{DnCNN,FFDNet}, compressed-sensing \cite{Lecouat2020Games},
super-resolution \cite{DnCNN}, and more. However, these deep neural
networks (DNN) and their building blocks are currently not well understood, making it ineffective to draw on knowledge from
decades of signal processing and optimization research for insight or
improvement. As a result, poorly understood tools such as
batch-normalization \cite{Ioffe2015}, residual-learning, and feature-domain processing, have become staples for deep-learning
architectures. So called ``unrolled optimization" DNNs \cite{unrolling} propose a promising avenue to remedy this by deriving their network architecture from $K$ iterations of an iterative optimization algorithm. Existing unrolled optimization DNNs are able to acheive \soa results on image restoration tasks without the use of the previously mentioned standard deep-learning toolbox  \cite{Simon2019, Lecouat2020Games}. However, these formulations mostly leverage their interpretability to reduce network parameter count in comparison to popular DNNs. In this work, we present a further desirable quality to be gained from interpretable model construction: \textit{robustness}. 

Our proposed method tackles natural-image denoising via unrolled convolutional dictionary learning (CDL), which we call CDLNet. The convolutional dictionary learning model is rooted in the sparse representation prior, where we assume our signals $\x \in \R^N$ may be represented by a linear combination of only a few vectors (atoms) from a larger dictionary, $\bm{D}$,
\begin{equation} \label{eqn:sparse_rep}
\exists ~ \z ~ \mathrm{s.t.} \quad \x = \bm{D}\z, \quad \norm{\z}_0 \ll N,
\end{equation}
where $\norm{\z}_0 \coloneqq \abs{\{i : \z[i] \neq 0 \}}$ is the $\ell_0$ pseudo-norm (a.k.a counting norm). With some assumptions on the level of sparsity in $\z$ and diversity (incoherence) of the dictionary atoms, stable and unique recovery of $\z$ from $\x$ may be guaranteed \cite{Mallat}. A primary focus of signal processing literature in the last decade and a half has been on the selection of a representation dictionary, in which learning based methods such as KSVD \cite{KSVD} brought \soa performance in image restoration tasks. Given a data-set $\{x_i\}_{i=1}^n$, the dictionary learning problem may be formulated with an $\ell_1$-norm relaxation of the $\ell_0$ pseudo-norm, which has been shown to promote sparsity,
\begin{equation} \label{eqn:dict_learn}
\underset{\bm{D}\in \mathcal{C}, \{\z_i\}}{\mathrm{minimize}} ~ \frac{1}{n} \sum_{i=1}^n \frac{1}{2}\norm{\bm{D}\z_i -
\x_i}_2^2 + \lambda \norm{\z_i}_1.
\end{equation}
The regularization parameter $\lambda$ gives a data-fidelity vs. sparsity trade-off, and $\bm{D}$ is restricted to the constraint set of columns in the unit-ball,
\begin{equation} \label{eqn:constraint}
\mathcal{C} = \{\bm{D} : \norm{\bm{d}^j}^2 \leq 1 \, \forall j\},
\end{equation}
to prevent arbitrary scaling of coefficients.
This formulation is amenable to the denoising problem, $\y = \x + \n$, by replacing the data fidelity term with in (\ref{eqn:dict_learn}) with $\norm{\bm{D}\z_i - \y_i}_2^2$. In this work, we consider the denoising problem with additive white Gaussian noise
(AWGN), where $\n \sim \mathcal{N}(0,\sigma_n^2\bm{I})$.

The problem in (\ref{eqn:dict_learn}) is non-convex and thus commonly solved via a Gauss-Seidel split, alternating between sparse-coding with a fixed dictionary, and updating the dictionary given sparse codes \cite{KSVD}. The iterative soft thresholding algorithm (ISTA) is one such sparse-coding algorithm, whose iterates are defined as,
\begin{equation} \label{eqn:ista}
\z^{(k+1)} \coloneqq \ST\left(\z^{(k)} - 
\eta^{(k)}\bm{D}^\top(\bm{D}\z^{(k)} - \y),\, \eta^{(k)}\lambda \right),
\end{equation}
where $\ST(\x,\tau)[i] \coloneqq \sign(\x[i])\max(0, \abs{\x[i]} - \tau)$ is the element-wise soft-thresholding operator with threshold $\tau \geq 0$, $\eta^{(k)}$ is a step-size parameter, and $\lambda$ is the sparsity penalty from (\ref{eqn:dict_learn}). In the case of a convolutional dictionary, we use $M$ filters $\bm{d}^j\in \R^{p}$ whose integer translates form the columns of $\bm{D}$. The application of the dictionary (synthesis convolution) is given by 
$\bm{D}\z = \sum_{j=1}^M \bm{d}^j \ast \z^j$, where $\z^j$ is the $j$-th channel of $\z$. 
The corresponding analysis convolution is defined channel-wise by $(\bm{D}^\top \x)^j = \overline{\bm{d}^j} \ast \x$ where $\overline{\bm{d}^j}$ denotes the reversal of the filter. In the case of 2D input signals $\x \in \R^{\sqrt{N}\times\sqrt{N}}$, we consider 2D square filters $\bm{d}^j \in \R^{\sqrt{p}\times\sqrt{p}}$. 

Making a connection between ISTA and DNNs is tempting due to their iterative use of linear operators followed by point-wise non-linearities. Our work builds off the seminal work of \cite{Gregor2010} by employing a learned ISTA encoder followed by a linear synthesis dictionary.

\section{Related Works}
The authors of \cite{Sreter2018} proposed an unrolled CDL DNN for image denoising and image inpainting, demonstrating results competitive with that of KSVD \cite{KSVD} in a fraction of the computational time. Our baseline proposed method builds off of their framework, introducing strided convolutions, preprocessing, and weight initialization as key steps to reaching denoising performance on par with \soa deep learning methods.

In \cite{Simon2019}, an argument is given suggesting the ill-conditioned nature of the convolutional sparse-coding (CSC) model in the representation of natural images. Their proposed model, CSCNet, uses large strides on the order of the filter size, along with averaging reconstructions from shifted input signals -- effectively returning to a patch-based approach. CDLNet demonstrates an alternative approach to CSCNet that is not inconsistent with their analysis, achieving superior performance on image denoising benchmarks without the use of ``shift-averaging".

The authors of \cite{Mohan2020} showed that several DNN denoising models, such as DnCNN \cite{DnCNN}, exhibit a catastrophic failure in denoising performance when presented with input noise-levels outside of the model's training range. They proposed ``bias-free" versions of the networks (ex. BF-DnCNN), which were demonstrated to posses the desired noise-level generalization property. Our work provides an alternate route for bringing robustness to DNN denoising by adapting the network's thresholds with the input noise-level. 

The authors of \cite{Isogawa2017} replace the ReLU non-linearities of DnCNN \cite{DnCNN} with soft-thresholding operators whose thresholds are proportional to the input signal standard-deviation. 
Our noise adaptive model instead employs soft-thresholding with thresholds proportional to the \textit{noise} standard-deviation, and is shown to generalize on noise-levels outside the training range.

In \cite{FFDNet}, the proposed model (FFDNet) requires an input ``noise-map" at inference, which allows it to handle spatially varying noise in addition to improved performance over a larger input noise-level training range. This noise statistic is only used as input to the network, whereas our method adapts the threshold of each layer of the model based on the estimated noise-level. 

\textbf{Summary of our contributions}: In this work we construct a convolutional dictionary learning network (CDLNet) that outperforms other CDL based DNNs while remaining in the CSC framework (Section \ref{sec:prop_arch}). From our interpretable model construction, we derive a noise adaptive model (Section \ref{sec:prop_ada}). Experimentally, we demonstrate that \soa performance in AWGN denoising is attainable when the model capacity is scaled to that of popular DNNs (Section \ref{sec:exp_single}). Additionally, we show that the proposed models with adaptive thresholds achieve \soa blind denoising performance and allow the model to generalize above and below training noise-levels (Section \ref{sec:exp_blind}). 

\section{Proposed Methodology} \label{sec:proposed}

\subsection{The CDLNet architecture} \label{sec:prop_arch}
\begin{figure*}[t]
    \centering
    \includegraphics[width=0.9\linewidth]{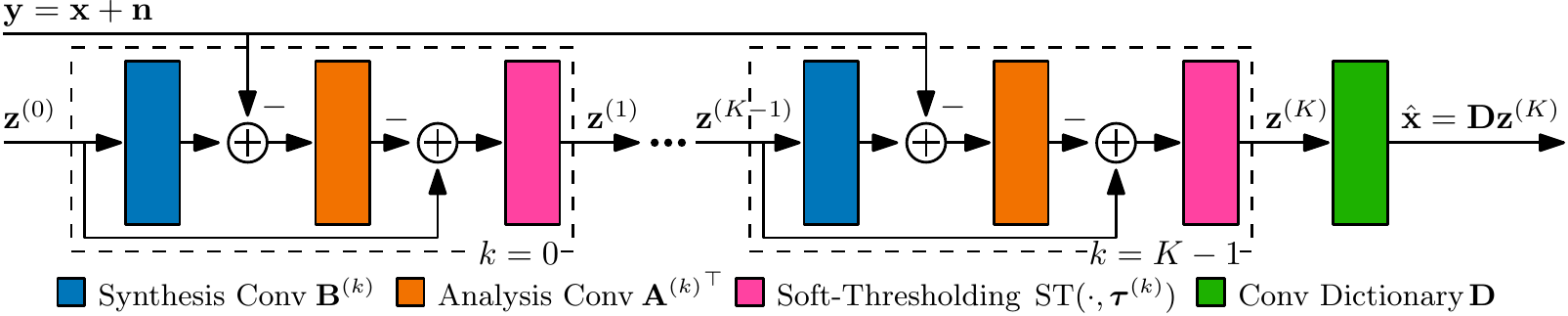}
    \caption{Block diagram of CDLNet. Analysis and synthesis convolutions map from $1$ to $M$ and $M$ to $1$ channels, respectively. We say that CDLNet does not process signals in a ``learned feature domain" to differentiate from the usage of multi-channel filtering ($M$ to $M$ channels) in DNNs such as DnCNN \cite{DnCNN}. Also note that CDLNet does not use batch-normalization or residual learning, in contrast to DnCNN \cite{DnCNN}.}
    \label{fig:block_diag}
\end{figure*}

In this section we introduce the convolutional dictionary learning network (CDLNet) for natural image denoising. The architecture involves a convolutional learned ISTA encoder with $K$ unrollings and a convolutional synthesis dictionary, $\bm{D}$,
\begin{gather} \label{eqn:CDLNet}
\hat{\x} = \bm{D}\z^{(K)}, \quad \z^{(0)} = \bm{0}, \quad k=0,1,\dots, K-1,\\
\z^{(k+1)} = \ST\left(\z^{(k)} - {\bm{A}^{(k)}}^\top (\bm{B}^{(k)}\z^{(k)} - \y), \, \boldsymbol{\tau}^{(k)} \right), \nonumber
\end{gather}
where $\Theta = \{[\bm{A}^{(k)}, \bm{B}^{(k)}, \boldsymbol{\tau}^{(k)}]_{k=0}^{K-1}, \bm{D}\}$ are the set of learned parameters. $\bm{B}, \bm{D}$ are all $M$-channel convolutional synthesis operators, and $\bm{A}^\top$ are $M$-channel convolutional analysis operators. The non-negative thresholds $\boldsymbol{\tau}^{(k)} \in \R^M$ are  subband dependent, corresponding to a learned weighted $\ell_1$ norm in (\ref{eqn:dict_learn}). The output denoised image is given by $\hat{\x} = \bm{D}\z^{(K)}$. The block diagram of the proposed CDLNet is given in Fig.~\ref{fig:block_diag}.

The convolution analysis and synthesis operators as shown are highly redundant transformations, increasing the number of coefficients from $N$ to $MN$ in analysis. This level of redundancy can become computationally burdensome as the number of filters ($M$) increases. To mitigate this, we can replace each of the convolution synthesis operators $\bm{A},\bm{B},~\text{or}~\bm{D}$, represented by $\bm{W}$, with a sub-sampled version $\bm{W}\Delta_s$, where $\Delta_s$ is the zero-filling operator with stride $s$. Note that $\Delta_s^\top$ is the sub-sampling operator with stride~$s$. For 2D images we consider $\Delta_s$ with stride~$s$ in both horizontal and vertical directions. The computational complexity of this network is roughly $O(KMNp/s^2)$ for input signals in $\R^{\sqrt{N}\times\sqrt{N}}$ and filters of size $\sqrt{p}\times\sqrt{p}$. Replacing the convolution synthesis operators $\bm{A},\bm{B},\bm{D}$  in (\ref{eqn:CDLNet}) with their sub-sampled versions allows for the possibility of learning more diverse filters while keeping the computation under control. 

\subsection{Noise-adaptive thresholds} \label{sec:prop_ada}
We leverage CDLNet's signal-processing and optimization derivation to provide adaptation to varying input noise-levels in a single model. We augment the learned thresholds in \eqref{eqn:CDLNet} to be proportional to the input noise standard deviation ($\sigma_n$), 
\begin{equation}
\boldsymbol{\tau}^{(k)} = \boldsymbol{\lambda}^{(k)}\sigma_n, \quad k=0,1,\dots, K-1 
\end{equation}
where $\boldsymbol{\lambda}^{(k)}\in\R^M$ is a learned parameter. We propose this augmentation of the thresholds based on the ``universal threshold theorem" \cite{Mallat}, from the Wavelet denoising literature,
$\tau = \sigma_n \sqrt{2\log_e N}$.
This can be shown to minimize the upper bound of the risk function of an element-wise denoising operator \cite{Mallat}.
With this augmentation, we interpret the proportionality constant $\boldsymbol{\lambda}^{(k)}_i$ as learning the gain factor between the image domain noise-level $\sigma_n$ and the $i$-th channel's noise-level at layer $k$. This framework has the added benefit of decoupling noise-level estimation from denoising, allowing for trade-off between accurate estimation and speed at inference time. We explore this trade-off using two different noise-level estimation algorithms at inference time, in Section \ref{sec:exp_blind}. We refer to CDLNet with the above noise-adaptive augmentation as CDLNet-A. 

\section{Experiments} \label{sec:exp}
\subsection{Training and Inference}
\textbf{Models}: are trained by the Adam \cite{adam} optimizer on the $\ell_2$-loss with parameter constraints,
\begin{equation}
\underset{
\substack{
\bm{A}^{(k)}, \bm{B}^{(k)}, \bm{D} \in \mathcal{C} \\ 
\boldsymbol{\tau}^{(k)} \geq 0
}
}{\mathrm{minimize}} \norm{\x - \hat{\x}(\y;\Theta)}_2^2,
\end{equation}
where $\mathcal{C}$ is the unit-ball constraint (\ref{eqn:constraint}) imposed on all convolution operators. Constraints are enforced by projection after each gradient descent step. Filters of size $7\times 7$ are used. We denote the models with $M$=32, and $K$=20, when trained on a single noise level, as CDLNet-S, and when trained over a noise range with and without adaptive thresholds as \mbox{CDLNet-A} and \mbox{CDLNet-B}, respectively. Similarly, we denote the models with $M$=169 and $K$=30 as Big-CDLNet. Unless specified otherwise, CDLNet models use stride 1 convolutions and Big-CDLNet models use a stride of 2. Our framework is implemented in PyTorch and provided online \cite{GithubLink}.

\textbf{Dataset}: All CDLNet models and variants are trained on the BSD432 dataset \cite{bsd}. Input signals are preprocessed with division by $255$, random crops of $128\times 128$, random flips and rotations, and mean-subtraction. Models trained across noise-levels are done so by uniform sampling of $\sigma_n \in \sigma_n^{\mathrm{train}}$.

\textbf{Training}: A mini-batch size of $10$ samples is used. An initial learning rate of $1e-3$ is used, and reduced by a factor of $0.95$ every $50$ epochs for a maximum of $6000$ epochs or until convergence. Similar to the method in \cite{Lecouat2020Games}, we backtrack our model to the nearest checkpoint upon divergence, reducing the learning rate by a factor of $0.8$. 

\textbf{Initialization}: We initialize all convolution operators with the same weights drawn from a standard normal distribution. This comes from the intuition that the majority of our learned channels will consist of band-pass signals modeling image texture, and so our filters should be zero-mean. Following \cite{Simon2019}, as an initialization step, we normalize the convolution operators by their spectral norm in correspondence with the maximum uniform step-size of ISTA.

\textbf{Noise-level estimation}: We employ two different noise-level estimation algorithms for blind denoising: one based on the median absolute deviation (MAD) of the input's diagonal wavelet coefficients \cite{Chang2000}, and the other based on the principal component analysis of a subset of the input's patches (PCA) \cite{Liu2013}. These estimators offer two ends of the trade-off between speed (MAD) and accuracy (PCA).

Experiments were conducted with an Intel Xeon(R) Platinum 8268 CPU at 2.90GHz, an Nvidia RTX 8000 GPU, and 4GB of RAM, running Linux version 3.10.0.

\subsection{Single noise-level performance} \label{sec:exp_single}
In these experiments, we train models for individual noise levels. Table \ref{table:single} shows the results of the CDLNet model in the small parameter count regime  (CDLNet-S) against the other leading CDL based DNN, CSCNet \cite{Simon2019}, and in the large parameter count regime (Big-CDLNet-S) against \soa denoising models DnCNN \cite{DnCNN} and FFDNet \cite{FFDNet}. CDLNet outperforms CSCNet in much less computational time at inference, without the use of stride, by using fewer filters and untying its weights between unrollings. We are also able to outperform the listed \soa deep learning methods by scaling our model (Big-CDLNet) to comparable size. Inference timings given show that our larger model's use of stride yields a manageable computational complexity. The non-learned method BM3D \cite{bm3d} is given as a classical baseline, and its timing is an order of magnitude greater than the learned methods.

\begin{table}[ht]
\centering
\caption{Denoising performance (PSNR) on BSD68 testset ($\sigma = \sigma_n^{\mathrm{train}} = \sigma_n^{\mathrm{test}}$). All learned models trained on BSD432. $\ast$~numbers reported in \cite{Lecouat2020Games}.}
\resizebox{\linewidth}{!}{%
\begin{tabular}{c|ccc|ccc} \hline
$\sigma$  & BM3D  & CSCNet$^\ast$ & CDLNet-S  & FFDNet & DnCNN & Big-CDLNet-S\\\hline
15        & 31.07 & 31.40  & 31.60            & 31.63  & 31.72 & {\bf 31.74}       \\
25        & 28.57 & 28.93  & 29.11            & 29.19  & 29.22 & {\bf 29.26}       \\
50        & 25.62 & 26.04  & 26.19            & 26.29  & 26.23 & {\bf 26.35}       \\\hline
Params    & -     & 64k    & 65k              & 486k   & 556k  & 510k        \\ 
GPU time  & -     & 143 ms & 9 ms            & 7 ms   & 23 ms & 15 ms       \\\hline
\end{tabular}%
}
\label{table:single}
\end{table}

In Table \ref{table:single}, we use stride 1 convolutions in our small model and stride 2 convolutions for our big model. Table \ref{table:stride} empirically verifies these choices as optimal by showing the effect of stride on output PSNR, averaged over a gray-scale version of the Kodak dataset \cite{Kodak}. Note that unlike CSCNet we do not employ any ``shift-averaging". We see that the redundancy of the large model allows for use of stride 2 without a denoising performance penalty.

\begin{table}[ht]
\centering
\caption{Effect of stride for $\sigma_n^{\mathrm{train}}=\sigma_n^{\mathrm{test}}=25$.\\ PSNR values averaged over gray-scale Kodak \cite{Kodak} dataset.}
\begin{tabular}{c|c|c|c|c}
\hline
Stride & 1 & 2 & 3 & 4 \\ \hline
CDLNet-S & {\bf 30.19} & 30.09 & 29.75 & 29.21 \\
Big-CDLNet-S & 30.37 & {\bf 30.39} & 30.28 & 29.83 \\ \hline
\end{tabular}%
\label{table:stride}
\end{table}

\subsection{Comparison of learned dictionaries}
\begin{figure}[ht]
\centering
\includegraphics[width=0.99\linewidth]{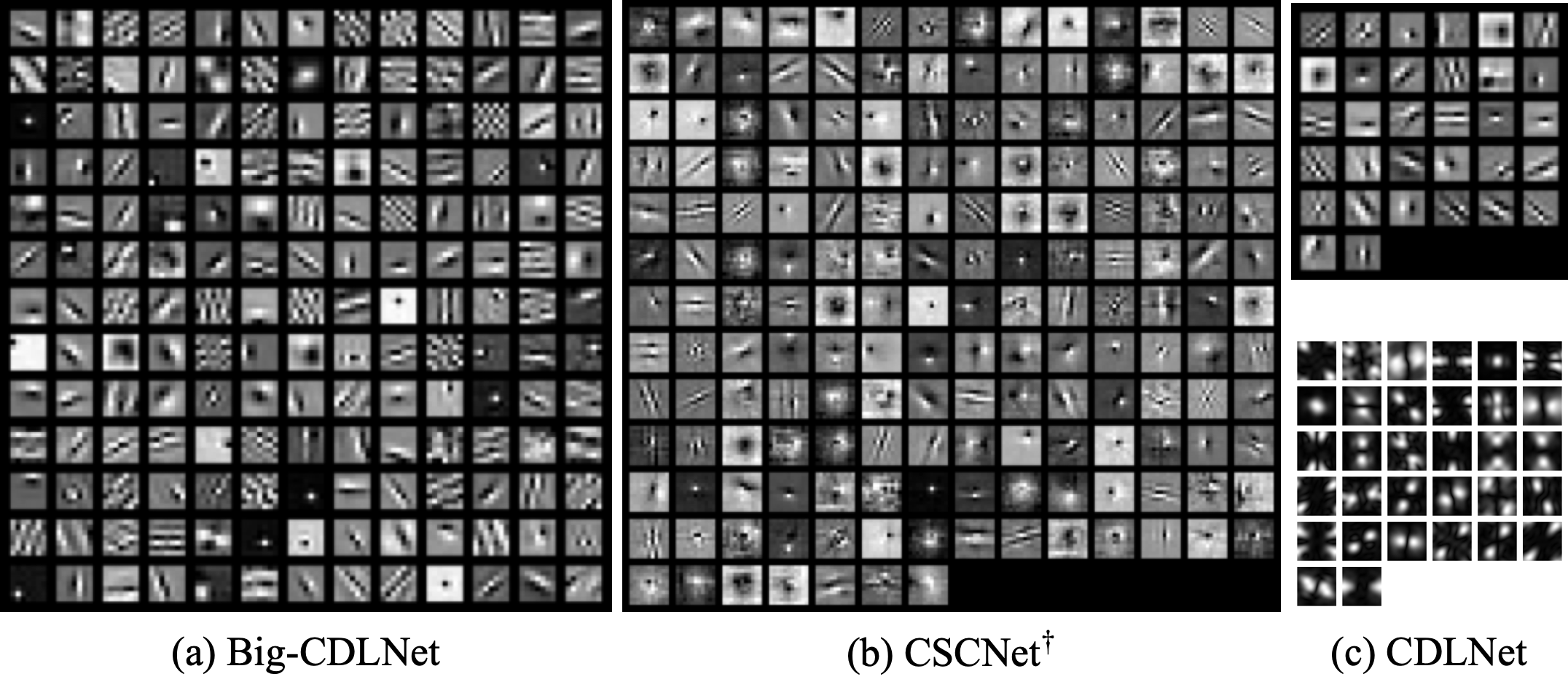}
\caption{Learned Filters for (a) Big-CDLNet, 169 filters of size $7\times 7$,  (b) CSCNet \cite{Simon2019}, 175 filters of size $11\times 11$, and (c) CDLNet (in spatial domain (top) and frequency domain (bottom)), 32 filters of size $7\times7$. $\dagger$ This figure is obtained from the models provided online by \cite{Simon2019}.}
\label{fig:filters}
\end{figure}

In this section we compare the learned representations from CDLNet and CSCNet \cite{Simon2019}. Fig. \ref{fig:filters} shows the convolutional dictionaries obtained from the CDLNet models and CSCNet. We see that Big-CDLNet offers a greater diversity in its learned filters compared to CSCNet. The baseline CDLNet model, despite having a relatively small number of filters, is shown to mostly learned directional ``Gabor-like" filters, with some texture components.

\subsection{Blind denoising and generalization across noise-levels} \label{sec:exp_blind}
\begin{figure}[ht]
\centering
\includegraphics[width=0.85\linewidth]{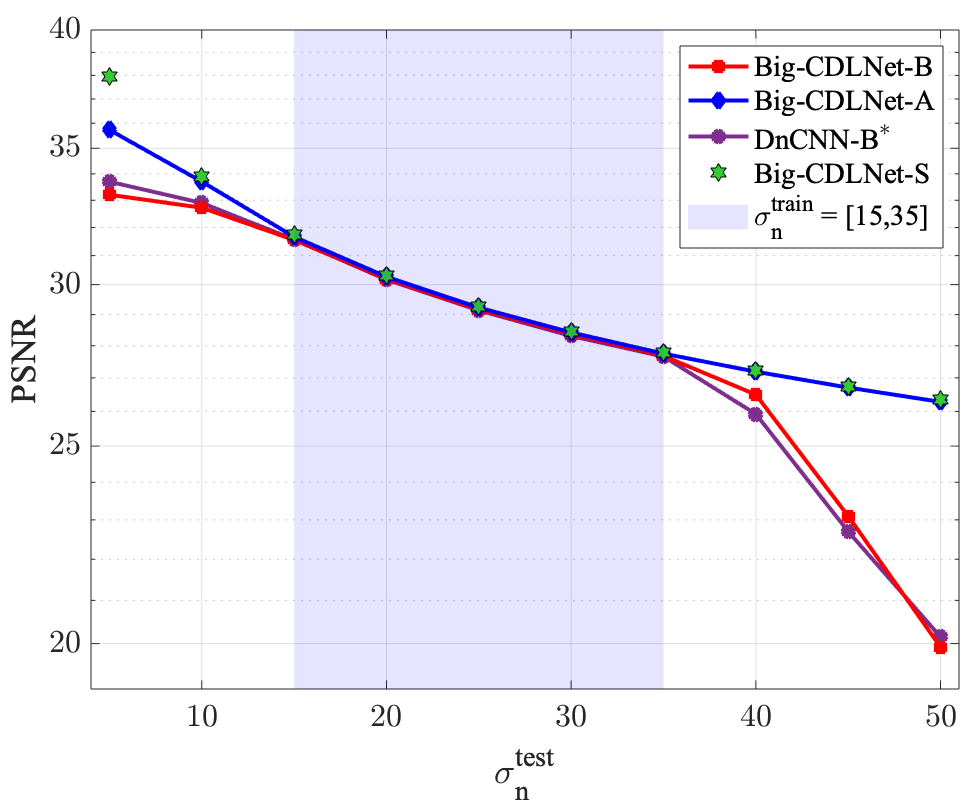}
\caption{Performance of different networks trained on $\sigma_n^{\mathrm{train}}=[15,35]$ and tested on different $\sigma_n^{\mathrm{test}}$. Average PSNR calculated over BSD68 \cite{bsd}.}
\label{fig:BlindPlot}
\end{figure}

In this section, we consider the blind denoising and generalization scenarios and compare the models equipped with the proposed adaptive threshold schemes to other models. In Fig. \ref{fig:BlindPlot}, we show the performance of the models trained on the noise range $\sigma_n^{\mathrm{train}}=[15,35]$ and tested on different noise levels $\sigma_n^{\mathrm{test}}\in[5,50]$. Big-CDLNet-A and Big-CDLNet-B refer to the proposed model trained on the training noise range with and without adaptive thresholds, respectively. Additionally, we trained the blind denoising version of DnCNN \cite{DnCNN}, DnCNN-B, on $\sigma_n^{\mathrm{train}}=[15,35]$, denoted as DnCNN-B$^\ast$ in Fig. \ref{fig:BlindPlot}. The single points on the plot in Fig. \ref{fig:BlindPlot} show the performance of Big-CDLNet-S model with single noise level training (i.e. $\sigma_n^{\mathrm{train}}=\sigma_n^{\mathrm{test}}$). 

As shown in Fig.\,\ref{fig:BlindPlot}, all networks perform closely over the training noise-range. On the other hand, when tested on noise-levels outside the training range, the network with adaptive thresholds (Big-CDLNet-A) greatly outperforms the other models. We observe that models without noise-adaptive thresholds have a very significant performance drop compared to the noise-adaptive model (Big-CDLNet-A) when generalizing above the training noise level, while Big-CDLNet-A nearly matches the performance of the models trained for a specific noise-level (Big-CDLNet-S) across the range. In spite of increasing input signal-to-noise ratio for noise-levels below the training range, we  observe that models without noise-adaptive thresholds have diminishing performance returns (note the plateau of Big-CDLNet-B and DnCNN-B$^{\ast}$ in $\sigma_n^{\mathrm{test}}=[5,15]$). On the other hand, denoising behavior of Big-CDLNet-A extends to the lower noise-range. We observe reduced generalization performance at the very low noise-level range of $\sigma_n = 5$. This may be explained by the need for a different thresholding model when the signal variance is much greater than that of the noise \cite{Mallat}.

\begin{figure}[ht]
\centering
\includegraphics[width=0.97\linewidth]{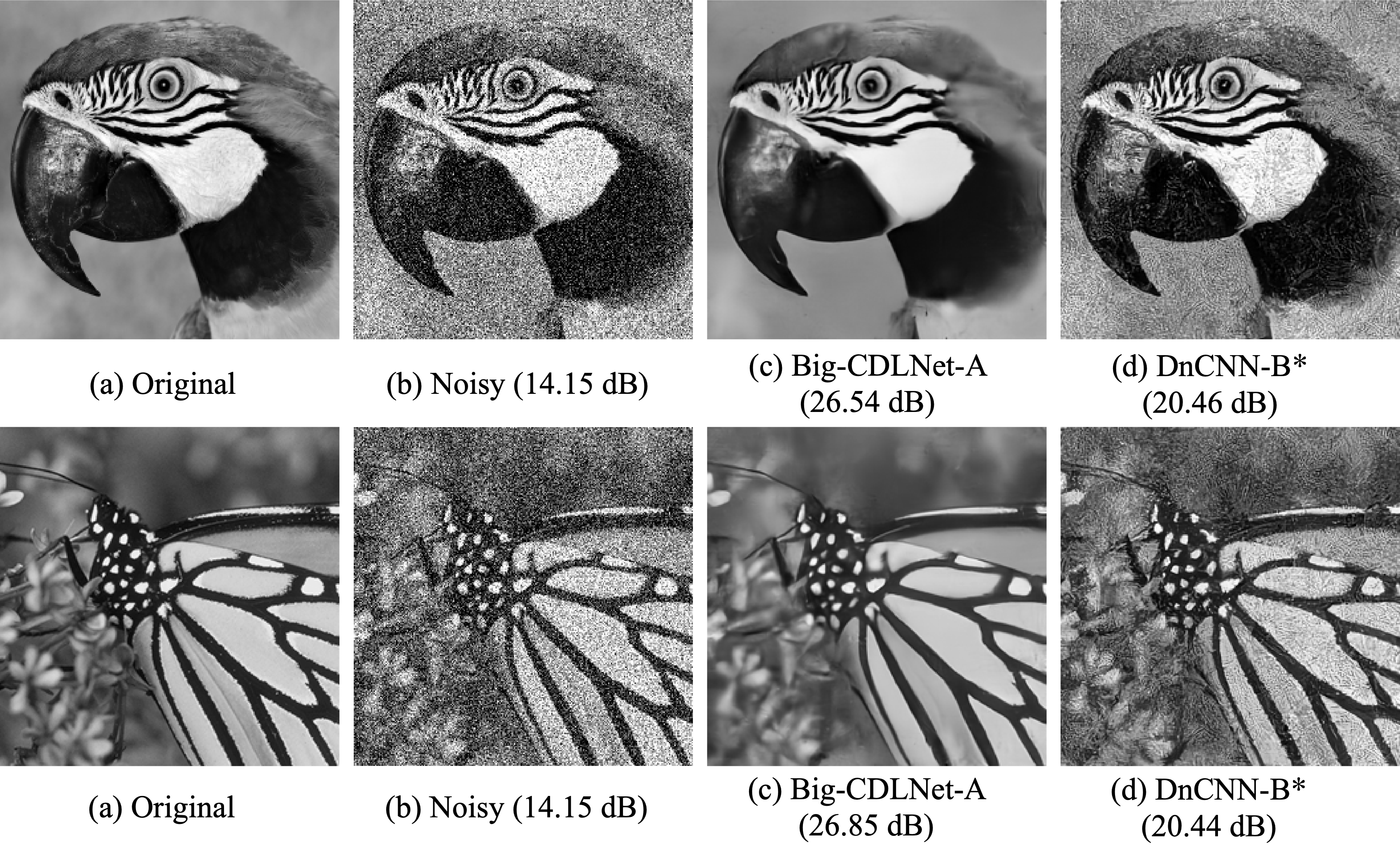}
\caption{Visual comparison of Big-CDLNet-A and DnCNN-B$^{\ast}$ trained on $\sigma_n^{\mathrm{train}} = [15,35]$ and tested on noise level $\sigma_n^{\mathrm{test}}=50$. PSNR value for each image is given in parentheses. Details are better visible by zooming.}
\label{fig:Parrot}
\end{figure}

A visual comparison of the denoising generalization of Big-CDLNet-A and DnCNN-B$^\ast$ is shown in Fig. \ref{fig:Parrot}. Our proposed adaptive model provides visually appealing results at the unseen noise-level ($\sigma_n^{\mathrm{test}}=50$), while DnCNN-B$^\ast$ fails to generalize and produces unwanted artifacts.

We further compare the blind denoising and generalization capabilities of the proposed method in Table \ref{table:generalization}. As observed in \cite{Mohan2020}, the BF-DnCNN model has reduced performance in the training range compared to DnCNN-B while avoiding the failure outside the training range. Big-CDLNet-A outperforms the DnCNN-B inside the training range and also provides improved generalization outside the range compared to BF-DnCNN. Note that the Big-CDLNet-A still has minor performance drop compared to a single noise level mode (Big-CLDNet-S) but this reduction is less significant than that of BF-DnCNN \cite{Mohan2020}. 

\begin{table}[ht]
\centering
\caption{Blind denoising and generalization comparison to DnCNN-B \cite{DnCNN} and BF-CNN \cite{Mohan2020}. All models are trained on $\sigma_n^{\mathrm{train}}=[0,55]$. Average test PSNR on BSD68 \cite{bsd} is reported.}
\resizebox{0.95\linewidth}{!}{%
\begin{tabular}{c|ccccc}\hline
\multirow{2}{*}{Model} & \multicolumn{5}{c}{$\sigma_n^{\mathrm{test}}$} \\
 & 5 & 15 & 25 & 50 & 75 \\ \hline
DnCNN-B & 37.65 & 31.60 & 29.15 & 26.22 & 18.74  \\
BF-DnCNN  & 37.72 & 31.58 & 29.12 & 26.17 & 24.63  \\
Big-CDLNet-A & \textbf{37.73} & \textbf{31.62} & \textbf{29.20} & \textbf{26.30} & \textbf{24.76}  \\ \hline
\end{tabular}%
}
\label{table:generalization}
\end{table}

In the blind denoising and generalization experiments of Table \ref{table:generalization}, Figures \ref{fig:BlindPlot} and \ref{fig:Parrot}, we employ the PCA based noise estimation algorithm \cite{Liu2013} at inference time. Table \ref{table:noise_est} shows the difference in denoising performance between using the ground-truth (GT)  noise-level and the two previously mentioned noise-level estimation algorithms. The PCA based algorithm allows us to attain near ground-truth denoising performance at the cost of increased computation. The wavelet based estimation method (MAD) offers essentially no computational overhead but significantly decreases denoising performance at the lower noise-level ranges.

\begin{table}[ht]
\centering
\caption{Effect of noise estimation algorithm on performance of the noise adaptive model Big-CDLNet-A.}
\resizebox{0.99\linewidth}{!}{%
\begin{tabular}{c|ccccl|c}\hline
\multirow{2}{*}{Est. Algo.} & \multicolumn{5}{c|}{$\sigma_n^{\mathrm{test}}$} & \multirow{2}{*}{GPU time} \\
 & 5 & 15 & 25 & 50 & 75 &  \\ \hline
GT & 37.75 & 31.63 & 29.20 & 26.31 & 24.80 & 13 ms \\
PCA \cite{Liu2013}& 37.73 & 31.62 & 29.20 & 26.30 & 24.76 & 23 ms \\
MAD \cite{Chang2000}& 37.18 & 31.55 & 29.18 & 26.30 & 24.75 & 13 ms \\\hline
\end{tabular}%
}
\label{table:noise_est}
\end{table}

\section{Conclusion}
We have proposed CDLNet as an interpretable deep neural network construction for the image denoising task. CDLNet achieves superior performance to other CDL based networks at a small parameter count, while avoiding the computationally expensive shift-averaging scheme. Additionally, we propose small-strided convolutions to enable the use of an increased model parameter count while retaining a manageable computational complexity. We show that such models perform competitively with \soa DNNs. Our results demonstrate that many of the popular tools from the deep learning toolbox (batch-norm, residual learning, feature domain processing) may not be necessary for the denoising task. We further leverage the interpretability of our network to propose a noise-level adaptive thresholding scheme that brings \soa blind denoising performance and robustness to noise-levels mismatch between training and inference.

\bibliographystyle{IEEEtrans}
\bibliography{bbl} 

\end{document}